\documentclass{optica-article}

\journal{opticajournal} 

\articletype{Research Article}

\usepackage{lineno}
\usepackage{chemformula} 
\usepackage[T1]{fontenc} 
\usepackage{siunitx}
\usepackage{booktabs}
\usepackage{multirow}
\usepackage{subfig}
\usepackage[]{graphicx}
\usepackage{MnSymbol,wasysym}
\usepackage{soul,xcolor}

\begin{document}

\title{Mode-Coupling-Driven Frequency Stabilization in Semiconductor Lasers with Bragg Grating Waveguide}

\author{M. R. Mahani\authormark{*}, Igor A. Nechepurenko\authormark{**}, Yasmin Rahimof\authormark{}, Andreas Wicht\authormark{}.}

\address{\authormark{}Ferdinand-Braun-Institut (FBH), Leibniz-Institut f\"{u}r H\"{o}chstfrequenztechnik, Gustav-Kirchhoff-Straße 4, 12489, Berlin, Germany}

\email{\authormark{*}Reza.Mahani@FBH-Berlin.de}
\email{\authormark{**}Igor.Nechepurenko@FBH-Berlin.de}

\begin{abstract*} 
Precisely stabilizing laser frequency is crucial for advancing laser technology and unlocking the full potential of various quantum technologies. Here, we propose a compact device for stabilizing frequency of a semiconductor laser through mode coupling effects, which provides enhanced sensitivity. Our proposed architecture features a main ridge waveguide with a Bragg grating, flanked by two curved ridge waveguides. This configuration exhibits an optical phenomenon characterized by a transmission crossing at the wavelength of the Bragg grating. Using particle swarm optimization strategy and employing efficient figures of merit, we achieve a high transmission crossing. The observed asymmetric transmission crossing not only holds the promise for an efficient and compact on-chip laser frequency stabilizer, but also fosters the development of novel sensing platforms with heightened sensitivity.\\
\end{abstract*}

\section{Introduction} 
Integrated photonics has emerged as a transformative field, revolutionizing the control of light on a nanoscale platform \cite{miller2009device, wang2020integrated, shen2017deep, miller2017attojoule, miller2010optical, jones2019heterogeneously, wang2020silicon, soref2010silicon, soref2018tutorial, hung2022near}. One of the most captivating aspects of this domain is the ability to engineer complex optical structures that offer unprecedented functionalities \cite{molesky2018inverse, cheben2018subwavelength, halir2018subwavelength, zheludev2012metamaterials}. In this context, the interplay between waveguide (WG) configurations and periodic structures, such as Bragg gratings (BGs), has led to advancements in signal processing \cite{burla2013integrated, shi2013silicon}, modulation \cite{stern2022large, ye2009numerical}, and sensing applications \cite{chen2019review}. The effective use of BGs in WG configurations for all-optical signal processing applications has been demonstrated \cite{sima2012all}. The modulation capabilities of BGs integrated with WGs, exhibiting their potential for high-speed data modulation has been shown \cite{pilipovich2002high}. Moreover, the sensing capabilities of these structures have been explored \cite{luan2020phase}, where BGs within WGs were employed for label-free biosensing with enhanced sensitivity \cite{luan2019label}.

Tilted BGs has recently attracted attention due to  their various applications which include (i) mode coupling and conversion allowing for various signal processing and modulation techniques \cite{wang2004compact,dai2012mode,yang2015widely}, (ii) sensitivity to external perturbations allowing for various sensing applications, including temperature and strain sensors and chemical or biological sensors \cite{gamal2015optical,saha2019highly,huang2023improving,butov2022tilted,chubchev2022machine,jean2024recent}, (iii) wavelength filtering and tuning allowing for precise control of the wavelengths of transmitted or reflected light \cite{hung2015narrowband}. 

These gratings, characterized by a tilt angle, offer control over mode coupling and propagation properties, rendering them indispensable in the realm of integrated photonics. By inducing controlled variations in the refractive index along the WG, tilted BGs enable efficient energy transfer between guided modes of different orders.

Stable laser sources are essential for various quantum optical experiments, including quantum communication, quantum computation, and quantum sensing \cite{duan2001long, aoki2021quantum, lezius2016space, becker2018space}. Here, we introduce an optimized on-chip frequency stabilizer design that leverages the phenomenon of asymmetric transmission crossing within tilted BG WGs. By utilizing this interplay, the device aims to provide a robust and compact laser stabilization solution for these quantum optical applications. The configuration consists of a primary narrow ridge WG comprising a tilted BG. Positioned on either side of the main WG are curved narrow ridge WGs, introducing an inherent asymmetry to the system. When light is launched into the main WG and its transmission is monitored in the flanking curved WGs, a pronounced transmission crossing is observed exactly at the Bragg resonance wavelength of the gratings. Through the integration of particle swarm optimization (PSO) algorithms and finite-difference time-domain (FDTD) simulation, we have enhanced this effect. 

This paper is structured as follows: Section \ref{sim} provides an overview of the theoretical underpinnings of the proposed device, the optimization algorithms used to maximize the effect, and a discussion of the results, focusing on the observed asymmetric transmission crossing and its implications. Finally, Section \ref{con} concludes the paper by highlighting the potential applications of this phenomenon in integrated photonics devices.

\section{Simulations and Optimization}
\label{sim}
\subsection{Structure}
\begin{figure}[htbp]
\centering
\includegraphics[scale=0.34]{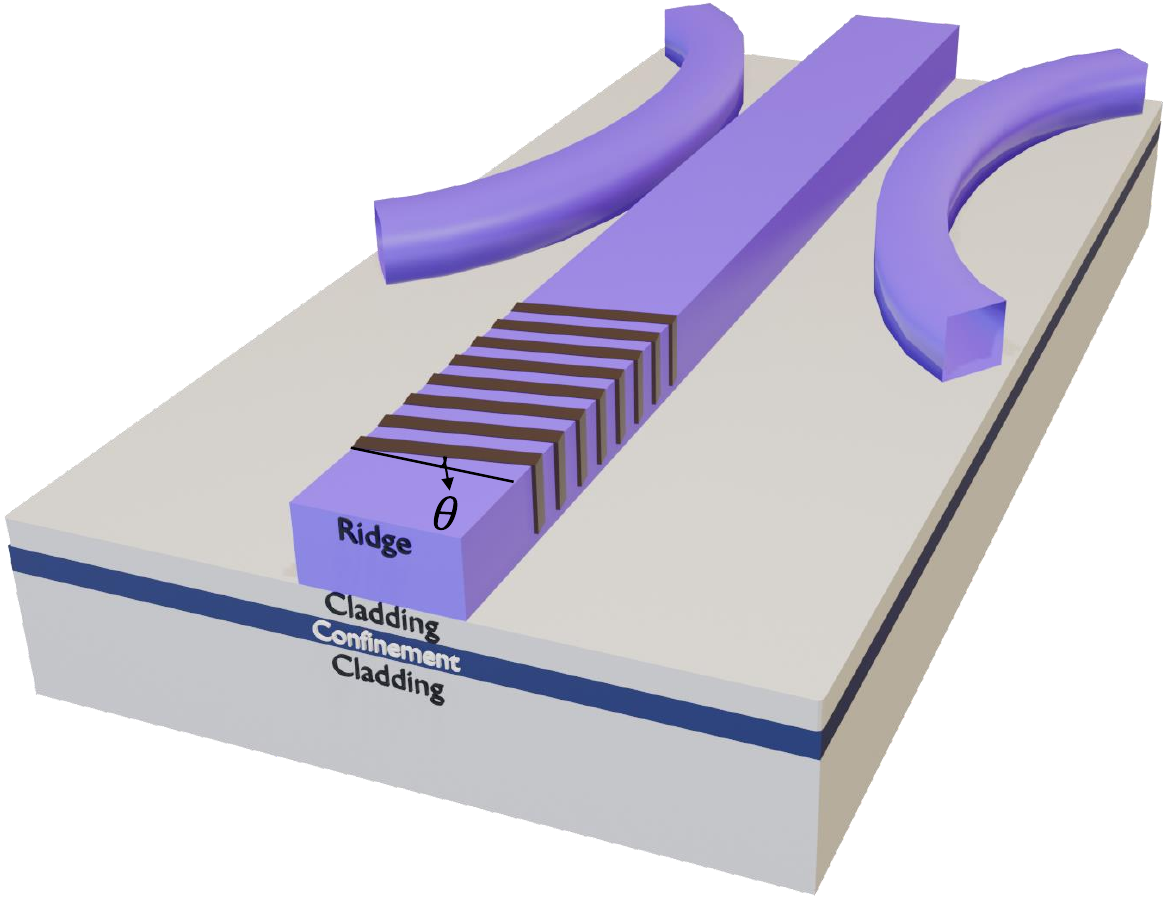}
\caption{ (Colour online) The 3D schematic structure of the device. The main ridge waveguide with a Bragg grating, flanked by two curved ridge waveguides.}
\label{fig:structure2}
\end{figure}

\begin{figure}[htbp]
\centering
\includegraphics[scale=0.4]{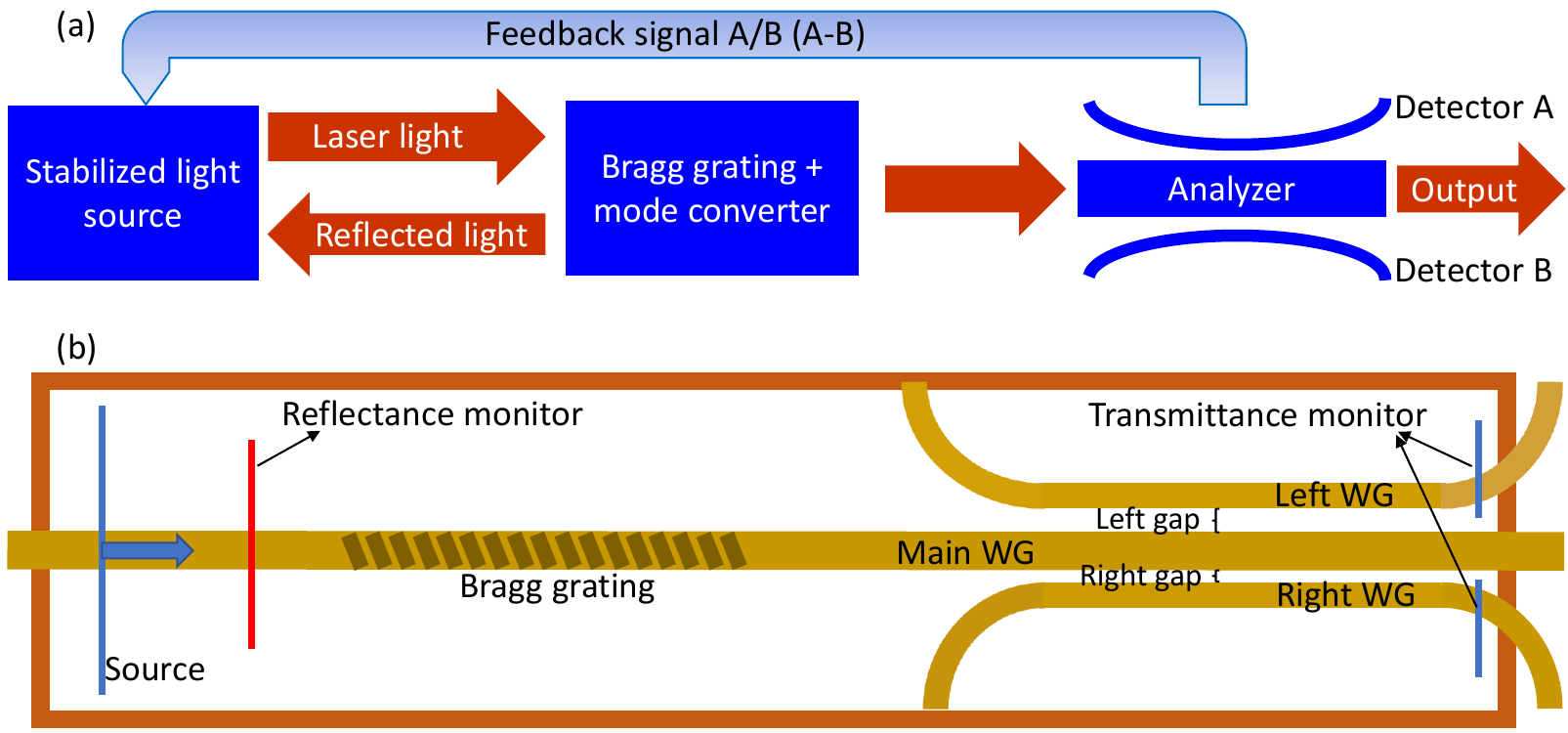}
\caption{ (Colour online) (a) The schematic of the device's concept, (b) full device, showing various parts; main WG, curved WGs, source, Bragg grating section, and the monitors.}
\label{fig:structure}
\end{figure}

The main narrow ridge WG, with a width of 250 nm and length of either 600 or 200 $\mu m$, serves as the backbone of the structure (Fig. \ref{fig:structure2}). The main WG comprises 72 tilted BG grooves with a period of 1 $\mu m$ and a refractive index of 3.5. These grooves are designed to enable frequency selective reflection at a Bragg wavelength. In addition, the tilt of the grooves allows an excitation of a higher order mode. A combination of a symmetrical fundamental mode and anti-symmetrical first order mode introduces an asymmetry to the field distribution in the main waveguide. There are also two curved WGs with different distance to the main WG, listed in the table \ref{tab:comp}. The different distance to the main WG imparts an asymmetry to the system. We refer to the one on the left side of the prorogation direction of light as the left WG and the other one as the right WG, and light is detected at the end of the side WGs. The schematic of the device's concept is shown in Fig.\ref{fig:structure}(a).
This concept is implemented and shown in Fig.\ref{fig:structure}(b), where various parts are introduced.

Our investigation focuses on the transmission characteristics of this integrated photonics device. Light is launched from one end of the main WG, adjacent to the BGs, and the transmitted light is monitored in the left and right WGs. The intensity of the signal in the right and left WGs is equal when the frequency of the incident light coincides with the frequency at which the transmission crossing occurs. As the frequency of incident light detunes, the coupling of modes in the main and side WGs changes. Depending on the sign of the change, the feedback signal can be created to tune the device back to the Bragg reflectance peak.
Using optimization described below, we have achieved a transmission crossing in the spectral domain, occurring precisely at the Bragg resonance wavelength of the gratings. 

\subsection{Optimization}
The simulation of the proposed integrated photonic device, was conducted using ANSYS Optics (Lumerical) software. The geometry of the device, including the main WG, tilted BGs and curved WGs was defined in the layout editor using Lumerical script. A simulation region encompassing the entire device structure was established. A broadband light source was defined in order to excite a fundamental mode in the desired spectral range for analysis.

Lumerical's FDTD solver was chosen for the simulation due to its computational efficiency \cite{wang2014precise, nechepurenko2023finite, mahani2023data, mahani2023designing}. To achieve accurate and reliable results the mesh for the simulation was 20 nm (wavelength/50). Perfectly matched layer (PML) was used to absorb scattered light. The simulation time was limited by an auto-shutoff value which was set to be $10^{-5}$ fraction of the energy left in the system. The reflectance was calculated for the fundamental mode by means of the mode projection \cite{snyder1983optical}. We used the built-in tools provided by ANSYS Optics simulation environment to collect and analyze the data.

The simulation procedure to optimize the proposed device was divided into two stages. The first set of optimization aimed to achieve 25\% reflectivity at the Bragg resonance (1063 nm) and the second set aimed to maximize transmission crossing ratio or difference. With such an approach, we split our optimization problem into two stages, each with smaller dimension. This is possible due to the low dependency of the Bragg reflectance on its tilt and the geometry of the side WGs. There is no dependency of the resonance wavelength on the tilting angle up to 15 degree and the amplitude of the resonance drops by approximately 10\% of the maximum. 

For the first optimization step, we choose the figure of merit (FOM) as the absolute value of the difference between the actual and desired reflectivity at 1063 nm. The optimization of the BG is rather straightforward, given a good analytical approximation for the period of the grating. The period is defined as $P= \frac{{N \lambda}}{2 {n_{eff}}}$, where $\lambda$ is the desired wavelength, $N$ is the order of the grating and $n_{eff}$ is an effective refractive index. The bandwidth of the Bragg resonance is very narrow, so the variation for the period should be of the order of 1\% from the analytical approximation. 
Refractive index variation determine the strength of the interaction of the WG mode with the grating. The number of the grooves also determines the strength of the reflection, and in general it can be fixed, when the refractive index is optimized. We implemented a PSO algorithm in connection with Lumerical software to optimize the period, the number of gratings and the refractive index. 

For the second set of optimization, to maximize the crossing ratio and difference at exactly Bragg resonance, we chose seven parameters and FOM that are described in the section \ref{sec_fom}. 
The set of seven parameters, which collectively define the device's geometry and structural characteristics, are:
BG tilt angle (1), width of the main WG (1), 
width of the side WGs (2), length of the side WGs (1),
distance between the left/right WGs with the main WG (we refer to these as left/right gaps) (2). The selection of the parameters is determined by the need to create an asymmetric coupling to the side WGs. Thus the geometry of the side WGs should be different. 

The PSO algorithm simulates the behavior of a swarm of particles in a multi-dimensional search space. Each particle, representing a potential solution with a set of parameter values, collectively explores the solution space to find the optimal configuration. The algorithm iteratively updates particle positions over multiple generations, where a generation refers to a specific iteration in the algorithm's optimization process. During each generation, the particle positions are updated based on their historical best and the swarm's best solution. The objective function is evaluated for each particle, and the algorithm converges towards the optimal parameter values. The optimization process continues until convergence criteria are met.

For the device optimization in the 7-dimensional parameter space, we used 50 particles (simulations) in each generation. We tried various generation sizes (number of particles in a generation). For larger size it requires many more simulation to converge, and smaller size is not sufficient to cover the entire parameter space efficiently. However, the results are not very sensitive to the exact number of the particles. The optimization procedure converges relatively quickly, after roughly 22 generations. We ran the simulations up to 27 generations ($27 \times 50 = 1350$ simulations) to ensure that the results are truly the optimum. We also repeated this procedure multiple times with different generation size, since the particles are generated randomly at the start of each optimization.

\subsection{Equations for the figure of merit}
\label{sec_fom}

The choice of the FOM is very important to obtain the desired outcome. For the second set of optimizations (optimizing transmission crossing ratio or difference), we had to try different FOMs to find the one that yields the best result.

For crossing ratio, we defined the following FOM (We refer to it as the first FOM) to evaluate the performance of each particle in PSO, quantifying the degree of asymmetric transmission crossing achieved by the device. The objective function takes into account the intensity imbalance between the left and right WGs at the Bragg resonance wavelength.
The parameters in Eq.\ref{firstfom} are: $\lambda_0 = 1.064 \times 10^{-6}$, $\lambda_1 = 1.0635 \times 10^{-6}$ and $\lambda_2 = 1.0645 \times 10^{-6}$.

\begin{equation}
\begin{split}
\text{FOM} = (\min(x) - 1) + (\min(x) > 1.05) \cdot \min(\{T_1(i_1), T_2(i_2)\}),\\
i_1 = \min\left(\left| \lambda - \lambda_2 \right|\right),
i_2 = \min\left(\left| \lambda - \lambda_1 \right|\right), \\
x = \left[\frac{T_1(i_1)}{T_2(i_1) + 1 \times 10^{-6}}, \frac{T_2(i_2)}{T_1(i_2) + 1 \times 10^{-6}}\right].
\end{split}
\label{firstfom}
\end{equation}

\begin{figure}
\centering
\subfloat[]{\includegraphics[scale=0.45]{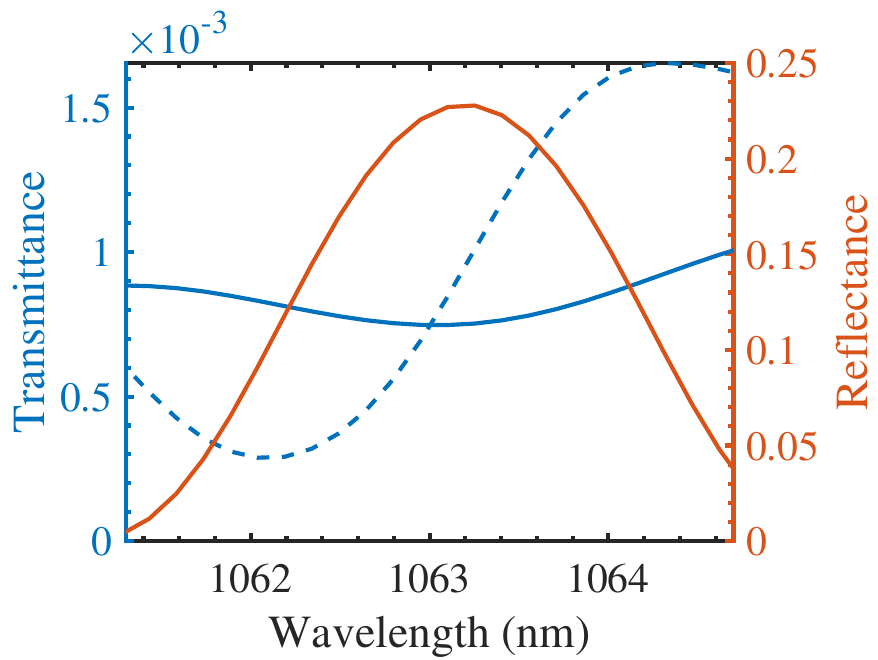}}
\subfloat[]{\includegraphics[scale=0.45]{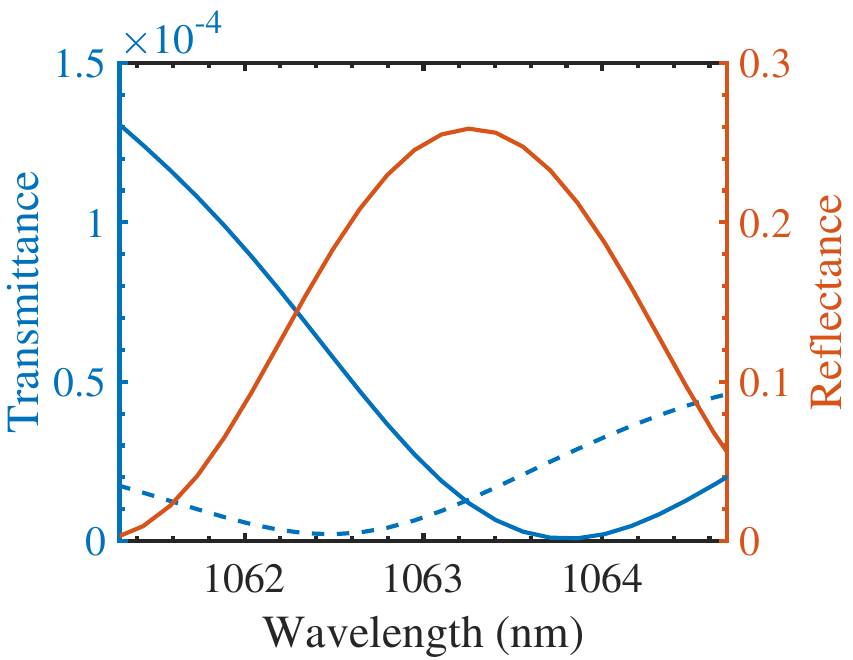}}
\caption{ (Colour online) Transmission crossing and Bragg resonance for the first FOM. (a) Main WG with the length of 600 $\mu$m (a crossing ratio of 2.9 and difference of 5.4$\times 10^{-4}$ can be achieved at 1nm away from Bragg resonance). (b) Main WG with the length of 200 $\mu$m (a crossing ratio of 16.8 and difference of 8.4$\times 10^{-5}$ can be achieved at 1nm away from Bragg resonance). The reflectance is shown on the right axis (red color). The transmittance in the left/right WG (solid/dashed blue line) are shown on the left axis.}
\label{fig:crossing}
\end{figure}

The goal was to maximize this imbalance, thereby maximizing the distinguishability of spectral shift.The range of the parameters and the optimized values are listed in the table \ref{tab:comp}, marked with the figure number, Figs.\ref{fig:crossing}(a) and Figs.\ref{fig:crossing}(b).

\begin{table}[htb]
 \centering \caption{Optimized parameters and their respective range.}

\begin{tabular}{|m{16em}|c|c|}
    \hline
Parameters for 1st optimization         & Parameter range       & Optimized parameters \\
    \hline
Period (P)                               & [0.99, 1.01] $\mu m$      &  1.0 $\mu m$  \\
Number of gratings (N)                   & [25, 72]                  &  72   \\ 
Refractive index (n)                     & [3.4, 3.5]                & 3.482   \\ 
    \hline
Parameters for 2nd optimization (main WG with the length of 600 $\mu$m)          &                           & Fig.\ref{fig:crossing}(a) - Fig.\ref{fig:cross2}(a)  \\
    \hline
Rotation angle ($\theta$) 			    & [0, 15] $^{\circ}$ 	     & 7.46 $^{\circ}$ - 0 $^{\circ}$ \\
Width of the groovs ($w_x$)	            & [80, 120] $nm$	         & 120 $nm$ -  120 $nm$  \\
Left/right WG length ($l$)          	& [10, 300] $\mu m$          & 267 $\mu m$- 271 $\mu m$\\
Left gap ($d_l$)                        & [50, 200] $nm$ 	         & 200 $nm$ -  66 $nm$ \\
Right gap ($d_r$)                       & [50, 200] $nm$ 	         & 94 $nm$ - 165 $nm$  \\
Width of the left WG ($w_l$ )           & [100, 400] $nm$ 	         & 368 $nm$ - 225 $nm$ \\
Width of the right WG ($w_r$)           & [100, 400] $nm$ 	         & 400 $nm$ - 300 $nm$ \\
    \hline
Parameters for 2nd optimization (main WG with the length of 200 $\mu$m)         &                            & Fig.\ref{fig:crossing}(b) - Fig.\ref{fig:cross2}(b)  \\
    \hline
Rotation angle ($\theta$) 			    & [0, 15] $^{\circ}$ 	     & 15 $^{\circ}$- 4.54 $^{\circ}$  \\
Width of the groovs ($w_x$)	            & [80, 120] $nm$	         & 110 $nm$  - 120 $nm$ \\
Left/right WG length ($l$)          	& [10, 30] $\mu m$          & 18.5 $\mu m$- 30 $\mu m$\\
Left gap ($d_l$)                        & [50, 200] $nm$ 	         & 59 $nm$  -  142 $nm$ \\
Right gap ($d_r$)                       & [50, 200] $nm$ 	         & 62 $nm$ - 50 $nm$  \\
Width of the left WG ($w_l$ )           & [100, 300] $nm$ 	         & 114 $nm$-  300 $nm$ \\
Width of the right WG ($w_r$)           & [100, 300] $nm$ 	         & 100 $nm$ - 226 $nm$ \\
    \hline
   \end{tabular}
   \label{tab:comp}
    \end{table}

The optimal structure demonstrates a large crossing ratio (16.8 for Fig.\ref{fig:crossing}(b)). The device has a large change of the signals detected between the left and right WGs. However, the overall signal is small in comparison with the transmitted light. In order to optimize for a higher transmission of light to the side WGs, we developed another FOM. The second FOM was used for Fig.\ref{fig:cross2}.

In this case, for crossing difference, we defined the FOM as the following. We first need to calculate the transmission difference $dT$ and its maximal variation $x_1$. 


\begin{equation}
\begin{split}
    dT &= T_1 - T_2, \\
    x_1 &= \min\left(-\min(dT), \max(dT)\right). \\
\end{split}
\label{secondfom1}
\end{equation}

Then we wrote a code that iterates through the elements of the $dT$ array to find points where the sign of adjacent elements changes. It keeps track of these positions in the crossings array. The code then checks if the crossing is greater than 0.01, it calculates FOM using a specific formula involving $i_0$ and x. Otherwise, it calculates FOM based on $x_1$.

\begin{equation}
\left\{ \begin{array}{l}
FOM = 10 \times \frac{1}{{\min \left( {\left| {{i_{cross}} - {i_0} + {{10}^{ - 3}}} \right|} \right)}},\,\,\,N > 0.\\
FOM = {x_1},\,\,\,N = 0.
\end{array} \right.
\label{secondfom2}
\end{equation}

In the equation above $N$ is the number of crossing in the given wavelength range, $i_{cross}$ is the index of the wavelength of the crossing closest to the index of the desired wavelength $i_0$. The optimized parameters and their range are listed in the table \ref{tab:comp}, marked with the figure number.

When we use this FOM, the optimized transmittance of light to side WGs becomes higher. The crossing also agrees well with the Bragg resonance (Fig.\ref{fig:cross2}). Such device is efficient to create a differential feedback signal.

\subsection{Discussions} 
The device architecture presented in this study, featuring tilted BGs integrated with curved WGs, has demonstrated an asymmetric transmission crossing at the Bragg resonance wavelength. By precisely tuning the parameters, the device can be used to select and filter specific spectral bands, a fundamental function in spectrometry or stabilizing laser frequency.

The transmission characteristics of the device are closely related to the spectral properties of the incident light. When light is launched into the main WG and monitored in the left/right WGs, the occurrence of a transmission crossing at the Bragg resonance wavelength signifies a specific wavelength's presence in the incident light. By measuring the wavelength at the crossing, the device can be used to analyze the spectral content of the input light.

\begin{figure}
\centering
\subfloat[]{\includegraphics[scale=0.45]{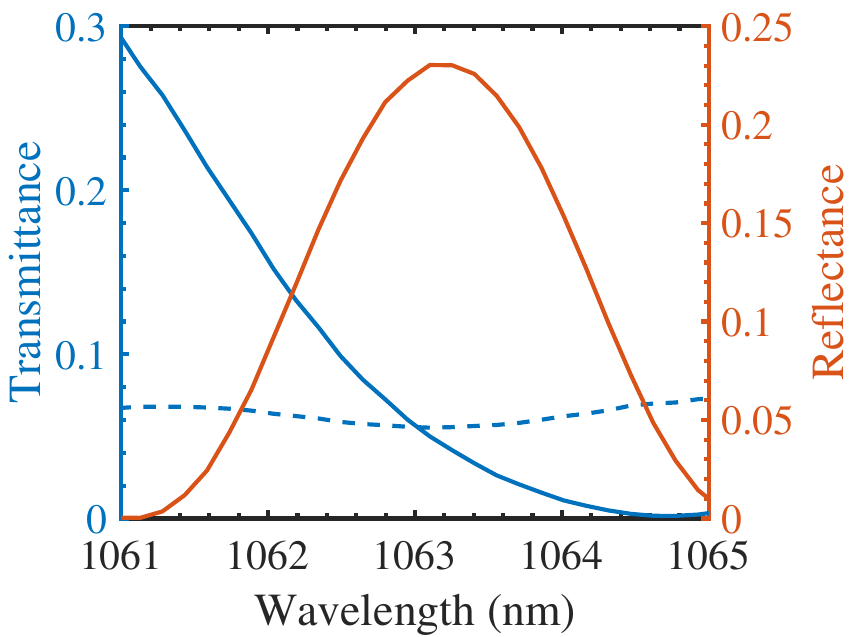}}
\subfloat[]{\includegraphics[scale=0.45]{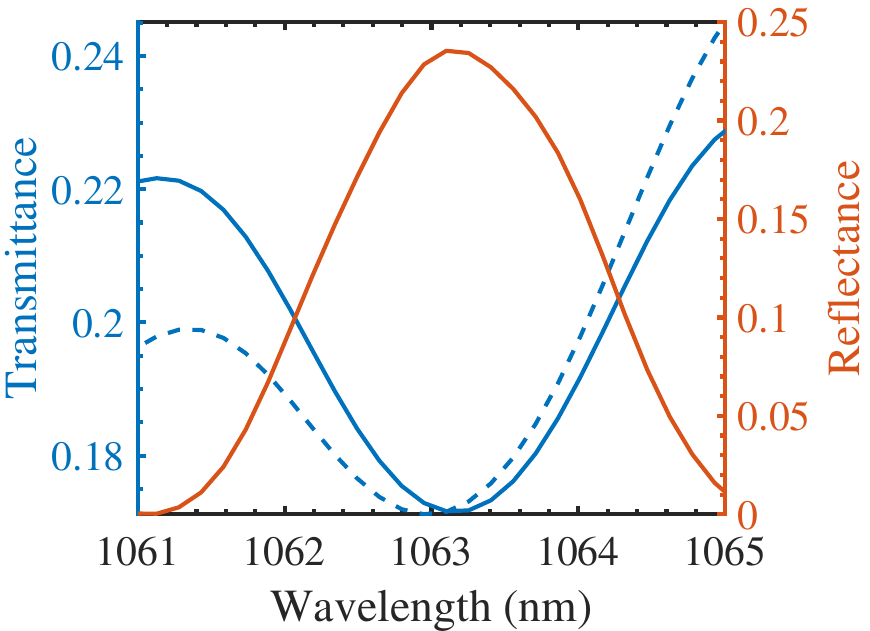}}
\caption{ (Colour online) Transmission crossing and Bragg resonance for the second FOM. (a) Main WG with the length of 600 $\mu$m (a crossing ratio of 5.7 and difference of 5.1$\times 10^{-2}$ can be achieved at 1nm away from Bragg resonance). (b) Main WG with the length of 200 $\mu$m (a crossing ratio of 1.1 and difference of 1.4$\times 10^{-2}$ can be achieved at 1nm away from Bragg resonance). The reflectance is shown on the right axis (red color). The transmittance in the left/right WG (solid/dashed blue line) are shown on the left axis.}
\label{fig:cross2}
\end{figure}

A high ratio of crossing transmission ($16.8$ in Fig. \ref{fig:crossing}(b)) suggests that the device can effectively identify specific wavelengths and is effective at separating wavelengths of interest from others. This can be important in applications where it is needed to distinguish and identify specific spectral lines in the presence of background noise.

On the other hand, if we want to increase the efficiency to create a differential feedback signal, we need to increase the transmittance of light in the side WGs. A weak signal ($10^{-4} - 10^{-3}$ in Fig. \ref{fig:crossing}) could be difficult to detect. While it might not be highly significant for quantifying intensity differences, it can still be useful for precise measurements in applications where the goal is to quantify subtle differences in spectral features. The small difference is important when we work with very faint spectral lines. However, this difference can be separately optimized using second FOM, as shown in Fig.\ref{fig:cross2}(a). Here the crossing difference has improved to reach 0.05 while the ratio is 5.7.

\begin{figure}
\centering
\includegraphics[scale=0.54]{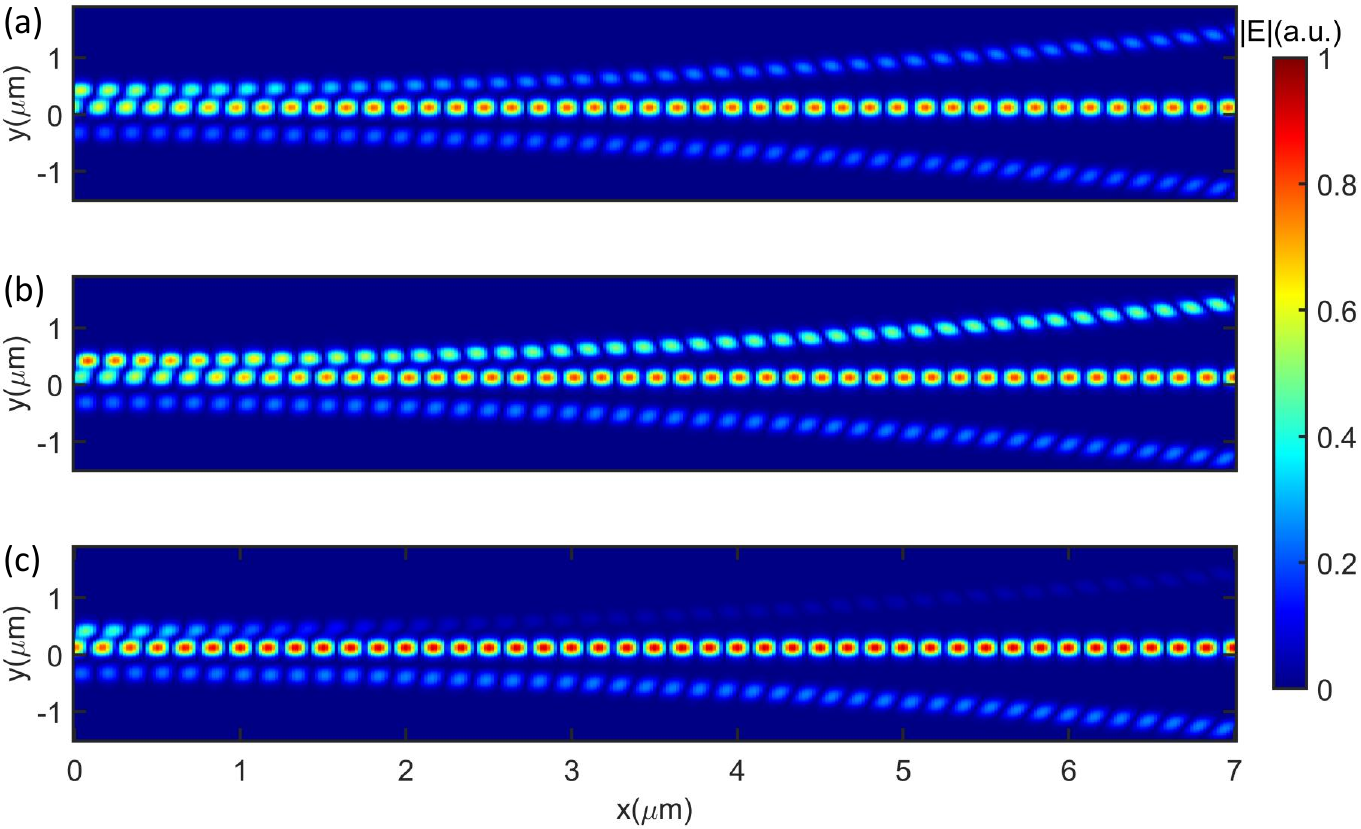}
\caption{ (Colour online) Electric field distribution in 2D space for a specific time (2000 fs after the light was launched) in the three WGs for three different wavelengths of the incident light. (a) $\lambda=1063$ nm, (b) $\lambda=1061.5$ nm, (c) $\lambda=1064.5$ nm.}
\label{fig:field}
\end{figure}

To gain an understanding of the device's behavior, we examine the electric field distribution in 2D space across the three WGs. We consider three cases, where the light pulse with the narrow spectra (few nms wide) is centered on three distinct wavelengths, Figs.\ref{fig:field}(a) $\lambda=1063$ nm, \ref{fig:field}(b) $\lambda=1061.5$, and \ref{fig:field}(c) $\lambda=1064.5$. If the incident light is centered on $\lambda=1063$, we observe the equal intensity in the side WGs, highlighting the transmission crossing at the BG resonance wavelength. At this wavelength, the transmission is equal in the side WGs, thus the intensity of the field is roughly the same (Fig.\ref{fig:field}(a)).

Further investigation reveals that as the incident wavelength deviates from the resonance point, the field distribution changes on the left WG. Figure \ref{fig:cross2}(a) illustrates that the transmission is notably higher in the left WG for shorter wavelengths ($\lambda=1061.5$). This asymmetry indicates the wavelength-dependent propagation of the electric field within the device (Fig.\ref{fig:field}(b)). Conversely, as the incident wavelength extends to longer values ($\lambda=1064.5$), a reversal in the transmission and respectively the electric field distribution occurs, with the right WG exhibiting a higher intensity (Fig.\ref{fig:field}(c)). We should keep in mind that the distance between the main WG and the left and right WG is different (66 nm and 165 nm, respectively). The coupling of the modes of the left WG with the modes of the central WG is much stronger than that of the right WG. The strong mode-coupling translates into short coupling length. This makes the system very sensitive to the optical length of the structure. As a result, the strong wavelength dependence of the transmission to the top WG is observed on Fig.\ref{fig:cross2}(a). On the contrary, the coupling of the modes of the right WG with the mode of the central WG is relatively weak, as a result the transmission of light to the right WG is almost constant.

The wavelength-dependent modulation in the electric field distribution across the WGs, in agreement with transmission crossing, elucidates the sensitivity of the proposed architecture, providing potential applications in laser frequency stabilization and sensing platforms. Since it is capable of filtering specific spectral bands, it can be used to isolate and stabilize a particular frequency. The setup's ability to detect specific wavelengths (high ratio of crossing transmission) is important for identifying the laser's emission wavelength, ensuring that it remains within the desired range. 

When the laser is operating at the wavelength close to the Bragg resonance wavelength (1063 nm), the device will exhibit a transmission crossing at this wavelength. This crossing is then used as a feedback signal. When the laser wavelength drifts away from the Bragg resonance, the transmission characteristics of the device will change. This change in transmission can be used as an error signal. Using a feedback control system, once the transmission characteristics change due to the drift, the control system brings it back to the the Bragg resonance wavelength. We would like to emphasize that, the setup's tunable grating parameters provide flexibility, allowing it to adapt to different laser sources and frequency stabilization requirements.

It is important to measure the sensitivity of the results with respect to the changes in the parameters.
In order to analyze the sensitivity of the proposed device, we varied the values of six most relevant parameters. The parameters are the left/right gap ($d_l$ and $d_r$), refractive index ($n$), length of the side WGs ($l$), width of the main WG and the tilting angle ($\theta$). We varied these parameters and compared the new transmission crossing wavelength $\lambda_{new}$ with the reference or optimized wavelength ($\lambda_{ref}$). The sensitivity is then calculated by taking the difference between transmission crossing wavelengths for $1\%$ change in the parameter values ($s=\lambda_{ref}-\lambda_{new}$). Calculated sensitivities for the left and right gap are $s=-0.0312$ nm and $s=0.1863$ nm,  respectively. The sensitivities for refractive index, length of the side WG, width of the main WG and the tilt angle are $s=0.0012$ nm , $s=-0.0046$ nm, $s=0.1149$ nm and $s=0.0001$ nm, respectively. The most sensitive parameters in our model are the the right gap and then the width of the main WG. The tilt angle seems to be the least sensitive parameter.

The device's performance may also be sensitive to environmental factors such as temperature fluctuations and vibrations. Ensuring stability in real-world conditions might be required. Fabrication imperfections and variations in the device's physical parameters (WG dimensions, grating properties) during manufacturing can also impact the device's performance. 

\section{Conclusion} 
\label{con}
The device we proposed here shows an asymmetric transmission crossing at the Bragg resonance wavelength. Through a combination of PSO algorithm and Lumerical software, we optimized this crossing at the Bragg resonance. We highlighted the details of our optimization approach and various FOMs and discussed the ones that gave rise to desired results and made the simulations converge fast.

Our investigation revealed the device's capacity to enhance spectral discrimination and sensitivity, leveraging the distinctive intensity imbalance at the asymmetric crossing. We obtained a high ratio of crossing transmission ($16.8$) and a large difference in transmitted intensities ($0.05$). We envisage applications in fields ranging from advanced optical stabilizing laser frequency to sensing and quantum applications.\\

\textbf{Author contributions:}
MRM, IN, YR, and AW conceived and planned the research. MRM, IN and YR carried out the simulations and optimizations. MRM, IN wrote the manuscript with input from all authors. All authors contributed to the interpretation of the results, provided feedback, and helped shape the research.
\\

\textbf{Research funding:} This work was supported by the VDI Technologiezentrum GmbH with funds provided by the Federal Ministry of Education and Research under grant no. 13N14906 and the DLR Space Administration with funds provided by the Federal Ministry for Economic Affairs and Climate Action (BMWK) under Grant No 50WK2272.\\

\textbf{Conflict of interest statement:} The authors declare no conflicts of interest regarding this article.\\

\textbf{Data Availability:} The data underlying this study are available from the corresponding author upon request.

\bibliography{References}
\end{document}